# Automated Droplet Size Distribution Measurements Using Digital Inline Holography

Santosh Kumar.S,[1] Cheng Li,[1] Chase E. Christen,[1] Christopher J. Hogan Jr.,[1] Steven A. Fredericks,[1,2] Jiarong Hong[1*]

[1]Department of Mechanical Engineering, University of Minnesota Twin Cities

[2]Winfield United, River Falls, WI

[*]To whom correspondence should be addressed: jhong@umn.edu

Droplet generation through spray breakup is an unsteady and non-linear process which produces a relatively dense, highly polydisperse aerosol containing non-spherical droplets with sizes spanning several orders of magnitude. Such variability in size and shape can lead to significant sources of error for conventional measurements based on laser scattering. Although direct imaging of droplets can potentially overcome these limitations, imaging suffers from a shallow depth of field as well as occlusions, which prevents the complete spray from being analyzed. In comparison, digital inline holography (DIH), a low cost coherent imaging technique, can enable high resolution imaging of the sample over an extended depth of field, typically several orders of magnitude larger than traditional imaging. In this study, we showcase an automated DIH imaging system for characterizing monodisperse and polydisperse aerosol droplet size and shape distributions in the 20 μm – 3 mm diameter range, over a large sample volume. The high accuracy of the technique is demonstrated by measurements of monodisperse droplets generated by a vibrating orifice droplet generator, achieving a resolution of ~14.2. Measurements of a polydisperse spray from a flat fan nozzle serve to establish the versatility of DIH in extracting a two-dimensional size-eccentricity distribution function, which indicates a strong semilogarithmic scaling between the two parameters that decays as the droplet migrates away from the nozzle. Due to its low cost and compact setup as well as high density of data obtained, DIH can serve as a promising approach for future aerosol characterization.

**Keywords:** Sprays; digital inline holography; droplet size distribution functions; size-shape joint PDF

## 1. Introduction

Liquid droplet generation via sprays is an aerosol process which is applied to efficiently transport liquids in gases in a range of fields (Lin & Reitz, 1998; Marmottant & Villermaux, 2004; Villermaux, 2007; Kooij *et al.*, 2018). The size and shape distribution functions of the droplets produced have a pronounced impact on droplet motion and evaporation rates, hence these distribution functions strongly affect the eventual output result of spray system application, including but not limited to influencing drug efficacy in spray based drug delivery (Labiris & Dolovich, 2003), combustion efficiency and soot formation in spray based fuel injection (Hayashi *et al.*, 2011), synthesized particle size in spray drying and pyrolysis (Nandiyanto & Okuyama, 2011; Zeng & Weber, 2014), coating morphology in spray deposition processes (Bhatia *et al.*, 2002; Fauchais *et al.*, 2014), and translocation and drift of crop protectants with agricultural sprays (Gil & Sinfort, 2005).

Measuring the droplet size distribution function in sprays is challenging when compared to most aerosols due to their high mass fractions (Tang *et al.*, 2018) and diameters that span multiple orders of magnitude in the supermicrometer size range. The two most common techniques for characterizing droplet size near the point of liquid atomization or breakup are laser diffraction (LD) and phase doppler particle analysis (PDPA). In LD, the light scattered by the droplets in the forward direction is detected on a radial sensor. Though it is possible to monitor monodisperse particle migration with LD (Jakobsen *et al.*, 2019), it is not an imaging-based technique (i.e. it is not single droplet sensitive), and hence its use requires a model based inversion to obtain the size distribution function of the detected droplets (Merkus, 2009). The use of model fitting enables the technique to successfully measure relatively dense aerosols generated by ultrasonic nebulizers (Kooij *et al.*, 2018; Kooij *et al.*, 2019). In comparison to LD, PDPA is a single droplet technique, in which the phase difference of scattered light at multiple angles is measured and inverted to obtain the size and velocity of individually detected droplets (Bachalo & Houser, 1984). It is presently the most common method to investigate fuel injection sprays (Soid & Zainal, 2011; Lee & Park, 2014). Though both are used widely, both LD and PDPA suffer from several limitations which can introduce error in droplet size distribution measurements. First, both require precise alignment of laser and detector, introducing significant complexity in implementation (Nuyttens *et al.*, 2007), and often require specifically designed experimental systems for measurements. The small sampling volume for PDPA (in the range of ~500 μm) introduces ambiguity when multiple droplets cross the volume, limits the largest droplets captured, and also requires long recording durations to ensure sufficient number of droplets are sampled (Soid & Zainal, 2011). Although LD records a much larger sampling volume in comparison, it suffers from a limited spatial resolution, and as already noted, provides only an ensemble size distribution function rather than single droplet analysis. More importantly, data analysis in both LD and PDPA is based on the assumption of spherical droplets, which is never strictly satisfied for a spray process and becomes problematic with increasing droplet size. Imaging-based techniques can overcome all the above limitations by providing high resolution size measurements on a per-drop basis over a large field-of-view from the captured images (Jiang *et al.*, 2010; Minov *et al.*, 2016). However, conventional imaging techniques have limited depth of field, i.e., the depth that the object appears to be in focus, which decreases drastically with increasing magnification. In addition, determining this depth of field precisely is challenging and thus leads to significant sampling uncertainties when measuring size distribution functions.

There is thus a need for improved droplet size distribution function analysis techniques for sprays. Digital inline holography (DIH) has recently emerged as an alternative imaging technique that captures three-dimensional (3D) information of objects with both high spatial resolution and an extended depth of field (typically more than 3 orders higher than traditional imaging) using a single camera (Xu *et al.*, 2001). DIH uses a coherent laser to illuminate a sample volume, while a camera captures the interference between the light scattered by an object in the sample and the unscattered part of the beam (referred to as a hologram hereafter). Once recorded, the hologram can then be numerically refocused by convolution with the Rayleigh-Sommerfeld diffraction kernel. Image processing enables determination of the 3D position, size, and shape of the object, as well as 3D velocity through tracking over time (Katz & Sheng, 2010). Information beyond what is measurable with LD and PDPA is hence measurable via DIH. Furthermore, the DIH measurement setup is considerably simpler (insensitive to misalignment of laser and camera) and less expensive than the setups for these conventional techniques (an order of magnitude difference in cost). The setup consists of a low power laser source (usable due to high efficiency of forward scattering light), a spatial filter, a collimation lens and a commercial camera with an imaging lens. The spatial filter increases the spatial coherence of the laser by focusing the light through a pinhole, which is then collimated into a parallel beam of light with the collimation lens. Any object placed in the path of the laser will produce a hologram, which is imaged on to the camera sensor by the imaging lens. Calibration is additionally facilitated via the use of a precision microruler, eliminating the need to use particle standards.

DIH has been recently demonstrated for the analysis of the motion and morphology of individual aerosol particles (Berg & Holler, 2016; David *et al.*, 2018; Giri *et al.*, 2019). DIH has also been used successfully employed in a diverse array of fields, including the study of social behaviors of flies (Kumar *et al.*, 2016), snowflake size distribution function and morphology (Nemes *et al.*, 2017), atmospheric mixing and cloud formation (Beals *et al.*, 2015), ocean sediment particle size and velocity transport (Graham & Nimmo Smith, 2010), oil droplets in oceans (Li *et al.*, 2017), bubbly wake behind a ventilated supercavity (Shao *et al.*, 2019), and preliminarily (without automated processing), droplets from sprays in compressible flow cross wind (Olinger *et al.*, 2014). Building upon these studies, in this work our goal is to develop an automated and rigorous data analysis approach for DIH to specifically determine the size distribution functions of spray generated droplets. The accuracy and the unique capabilities of DIH for droplet characterization are demonstrated through the measurements of monodisperse vibrating orifice generated droplets (Berglund & Liu, 1973) as well as those produced in a polydisperse agricultural spray in a low speed wind tunnel. Not only do we utilize DIH to infer the one dimensional size distribution function, but also a two dimensional size-shape distribution function (diameter and eccentricity), as both spherical and non-spherical droplets are easily analyzed. With the developed data and demonstrated data analysis approach, we anticipate that DIH can be utilized for spray droplet size distribution characterization with regularity.

## 2. Experimental Methods

We apply DIH to examine the size distribution functions of droplets in the 20 μm – 3 mm diameter range. Though these are on the larger end of what are typically considered aerosol particles or droplets, droplets in this size range are specifically chosen both with the target application of agricultural sprays in mind and to demonstrate the ability to analyze non-spherical droplets (which are often larger in size). DIH can be easily extended to examine droplets down to 1 μm in diameter, and to droplets which are submicrometer in dimension with appropriate increase in the imaging magnification and a decrease of the laser wavelength. The following sub-sections describe the DIH system, examination of monodisperse droplets, and DIH measurements in a recirculating wind tunnel housing an agricultural spray system.

*2.1. Digital Inline Holography*

Figure 1 illustrates a schematic of the DIH system across a recirculating wind tunnel (designed by Hambleton Instruments, Hudson, WI) capable of achieving wind speeds of 8 m/s. The test section, located opposite to the fan section, is 0.91 m in width, 1.83 m high, and 3.20 m in length, and is equipped with a spray nozzle mount on a one dimensional traverse (capable of moving vertically from the top to the bottom using a rate controlled stepper motor). Optical access is provided for measurements across the test section, and a mist eliminator located at the downstream end of the spraying section prevents the liquid from being recirculated into the tunnel.

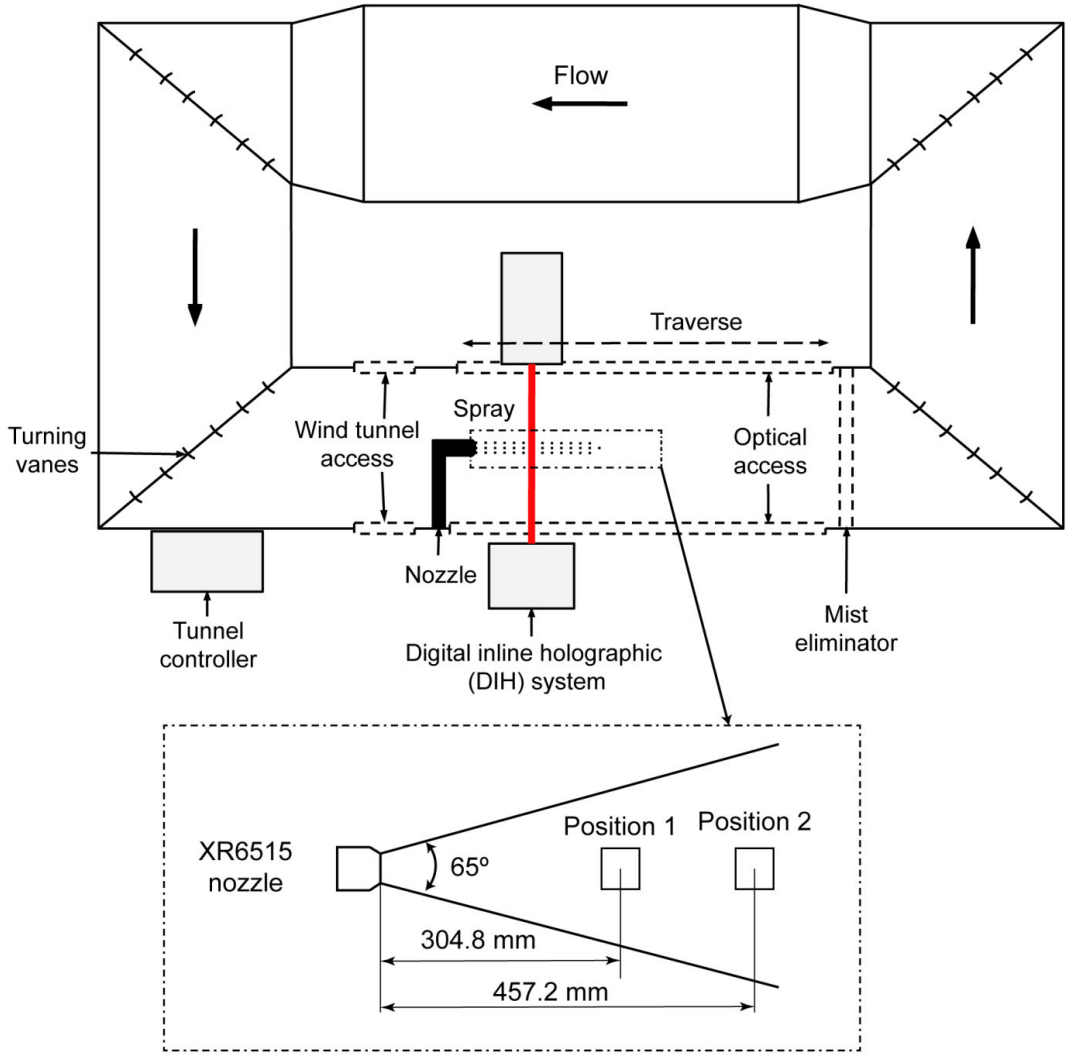

**Figure 1.** A schematic diagram of the recirculating wind tunnel where measurements were made, indicating the flow direction, location of the nozzle, the spray position and the position of the digital inline holographic imaging system. The lower inset qualitatively marks the spray fan from an XR6515 nozzle and the two imaging positions along the centerline.

A photograph of the DIH system across the wind tunnel optical access test section is provided in Figure 2. It consists of two parts, the illumination system (right side of Figure 2) and the imaging system (left side of Figure 2). The illumination system includes a 12 mW helium-neon laser (REO Inc.), a neutral density (ND) filter to control the laser intensity, a spatial filter (Newport Inc.) to increase laser spatial coherence, and a collimation lens with 75 mm focal length (Thorlabs Inc.) all mounted on an optical breadboard. Collectively these optical components yield a 50 mm diameter collimated Gaussian laser beam, which is aligned and pointing at the imaging system using two turning mirrors. The imaging system consists of a high speed camera (Phantom v710), an imaging lens (Nikon 105 mm f/2.8) and a computer to control the camera. The entire DIH system is mounted on a traverse which can be moved to different positions along the direction of flow in the wind tunnel. Taken along with the vertical spray traverse, this facility enables us to investigate specific locations within the spray fan. To calibrate the DIH system, we image a precision microruler (Thorlabs Inc.) to obtain a spatial resolution of 18.2 μm/pixel (see supporting information for additional details), which, as shown subsequently, is additionally confirmed by the monodisperse droplet measurements.

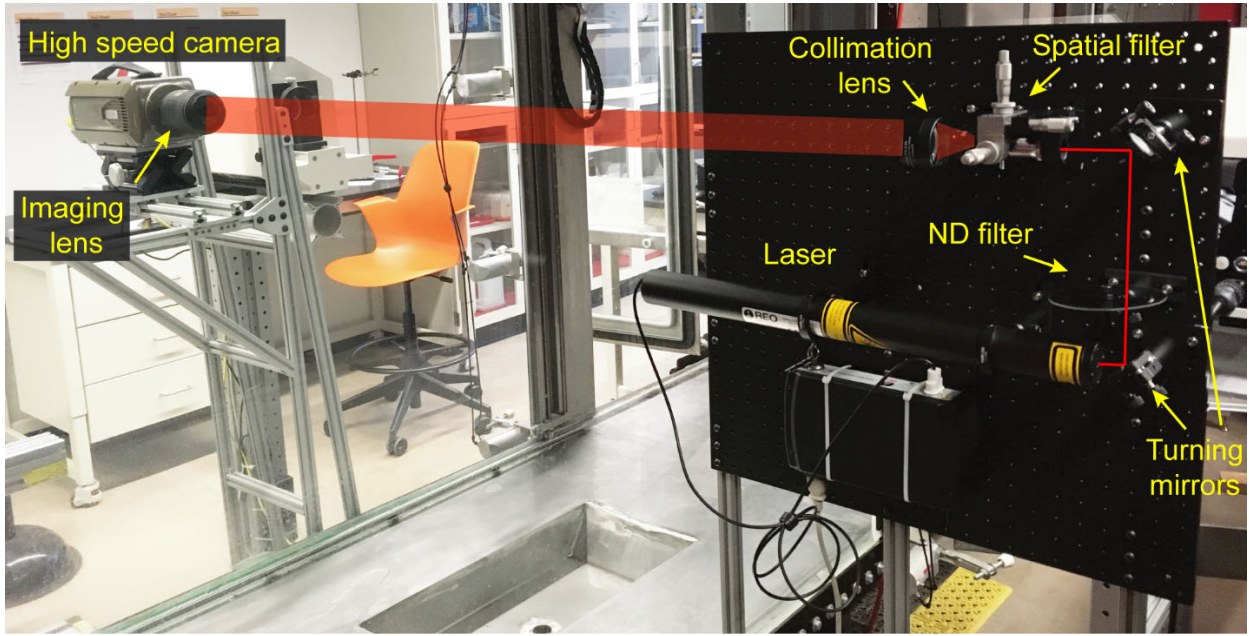

**Figure 2.** A photograph of the DIH imaging setup straddling both sides of the optical access region of the wind tunnel.

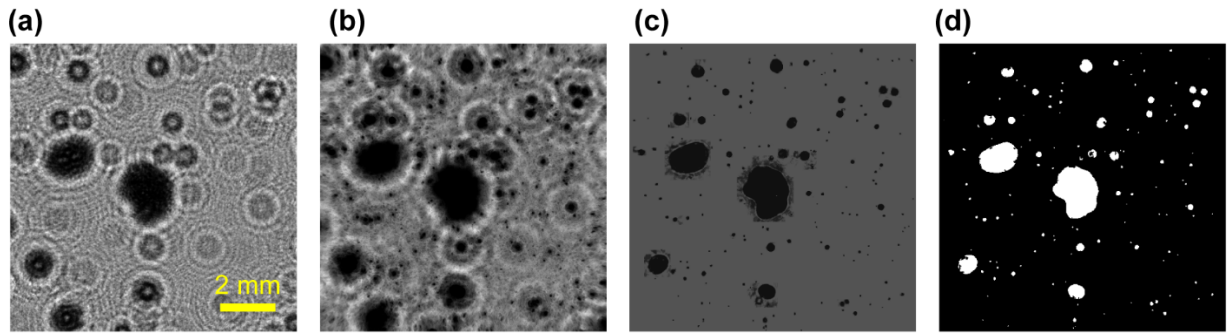

**Figure 3.** Demonstration of the steps of our automatic droplet detection and sizing algorithm, illustrated using a sample image: **(a)** the original hologram, **(b)** the reconstructed combined minimum intensity image, **(c)** the reconstructed image after refocussing of each droplet, **(d)** segmented binary image.

For all measurements reported, holograms were collected via the imaging system by using the camera to record either a 4.66 mm × 4.66 mm (monodisperse test case) or a 9.32 mm × 9.32 mm (polydisperse test case) field-of-view with 2.5 μs exposure time, respectively, under continuous laser illumination (see Figure 3a for a sample hologram). The captured holograms were processed using an automated MATLAB code to extract the droplet position, size, and shape. The main steps of the processing algorithm are preprocessing, reconstruction, in-focus image identification, binary image segmentation, and feature extraction. In preprocessing, a time averaged background image was subtracted from each hologram in order to enhance the fringe contrast and eliminate any stationary artifacts due to dust or dirt in the optical path, resulting in the background corrected hologram. Next, the hologram was reconstructed by convolution with the Rayleigh-Sommerfeld diffraction kernel to obtain the 3D intensity field containing all of the droplets (Figure 3c shows a minimum intensity projection of this field). Subsequently, we implemented the hybrid algorithm described by Gao *et al.* (2014) to automatically segment the minimum intensity projection to identify the droplets, and further refined their position in the longitudinal direction (along the laser beam) using a sharpness based focus metric (Gao *et al.*, 2014). Comparing the result of the refinement (Figure 3c) with the projected image, we can clearly observe the sharp edges that help accurately measure the size and shape of the each droplet. The in-focus grayscale image is finally binarized using an iterative thresholding algorithm performed on a local region of the image around each droplet, which identifies a mean intensity value that separates the droplet in the foreground from the background. A watershed segmentation step is also used on the binary image to separate occluding droplets (Figure 3d). As the sharpness based focus metric depends on the use of intensity gradients, we limit the aforementioned approach to droplets that are at least 100 pixels in area and implement an intensity based focus metric to locate the smaller droplets (Dubois *et al.*, 2006). Once the droplets are segmented, we extract their equivalent diameters, eccentricities, and positions, with details provided in the subsequent *Results & Discussion* section. We have included a supplementary video (supplementary video 1) which illustrates a side-by-side comparison of all the steps of the algorithm from the hologram to the final segmented objects for the monodisperse droplets.

*2.2 Monodisperse Calibration Measurements*

The accuracy of our DIH approach is validated through measurements of monodisperse droplets, which demonstrates DIH's ability to resolve narrow size distributions, similar to mobility analyzer resolution measurements with monomobile standards (Ude & de la Mora, 2005). Monodisperse droplets were generated using a vibrating orifice aerosol generator (VOAG, Model 3450, TSI Inc.), using 0.025% dioctyl sebacate in isopropyl alcohol. The nozzle head of the vibrating orifice aerosol generator with a 50 μm diameter orifice was mounted vertically above the DIH measurement volume, pointing downward with a dispersion air flow rate of 1.5 l/min. With the wind tunnel flow off, the generated droplets were allowed to migrate into the DIH measurement volume via gravitational settling. The VOAG was operated with a liquid feed rate of 1.003 ml/min, and a frequency of 14.98 kHz, which were verified using a mass calibration and an external oscilloscope, respectively. These operating parameters yielded a nominal droplet diameter of 128.6 μm (Berglund & Liu, 1973). For size distribution function measurements, we collected images for a total duration of 5.2 s at 25000 frames/s, yielding ~130000 holograms, out of which ~2600 images are subsampled to ensure each droplet was not counted more than once.

*2.3 Recirculating Wind Tunnel Spray Measurements*

The versatility of DIH measurements in characterizing the size and shape of droplets (specifically a size dependent shape) is showcased through measurements of a water spray (tap water at 19 ºC) from an XR6515 flat fan nozzle at a pressure of 152 kPa (measured across the nozzle). Holograms were captured at two positions along the centerline of the spray at distances of 304.8 mm (74.3$D_N$) and 457.2 mm (111.5$D_N$) downstream of the nozzle exit, where $D_N$ corresponds to the width of the nozzle orifice. The experiments were performed with a mean tunnel flow speed of 4 m/s and the spray aligned parallel to the flow direction. A sampling rate of 500 frames/s ensured that no droplets were imaged more than once over the same field of view. Images were collected for a duration of 1.6 minutes, resulting in ~49000 holograms. During the experiment, we monitor the relative humidity and temperature in the wind tunnel test section to be ~80% and 30 °C, respectively.

## 3. Results & Discussion

### 3.1. Determination of Size and Eccentricity Distribution Functions

After image processing, DIH enables automatic determination of droplet dimensions, a large fraction of which are not spherical (see Figure 3 & supporting information). While this is problematic in LD and PDPA data interpretation (due to the spherical droplet assumption), DIH facilitates quantification of both the size and extent of asphericity for such droplets. First, for each droplet, the diameter ($D$) is calculated as the projected area equivalent diameter, $(4A_\mathrm{p}/\pi)^{1/2}$, for the droplet of projected area $A_\mathrm{p}$. Second, the major ($b$) and minor ($a$) axes are identified by an ellipse fit on the segmented 2D droplet. These geometric parameters are then used to compute the eccentricity, $e = \sqrt{(1-(a^2/b^2))}$, which spans from 0 (for a perfect circle) to 1.

Each measurement yields the diameter and eccentricity of $N$ droplets. For VOAG generated droplets, we simply elect to construct a number based probability distribution function (PDF) to examine DIH performance on linearly spaced bins. For polydisperse flat fan spray droplets, number based size distributions $\frac{dN}{d(\log D)}$ are determined by selecting 20, 30, 40, or 50 logarithmically spaced bins in the 10-3000 μm range, placing each droplet in the appropriate bin, and then normalizing the bin value by the logarithmic bin width. The cumulative size distribution function at diameter $D_j$ is computed as:

$$\text{Number CDF} = \frac{\sum_{i=1}^{i=j}\left(\frac{dN}{d(\log D)}\right)_i \delta(\log D)_i}{\sum_{i=1}^{i=i_\max}\left(\frac{dN}{d(\log D)}\right)_i \delta(\log D)_i} = \frac{1}{N}\sum_{i=1}^{i=j}\left(\frac{dN}{d(\log D)}\right)_i \delta(\log D)_i \quad (1)$$

where $i_\max$ is the total number of diameter bins, $\left(\frac{dN}{d(\log D)}\right)_i$ is the size distribution function value, $\delta(\log D)_i$ is the width of logarithmic bin $i$ and $j$ denotes the bin with upper limit diameter $D_j$. Volume based size distribution functions, $\frac{dV}{d(\log D)}$, are calculated from the number based size distribution function via the equation:

$$\left(\frac{dV}{d(\log D)}\right)_i = \frac{\pi}{6}\left(\frac{dN}{d(\log D)}\right)_i D_i^3 \quad (2)$$

where $D_i$ is the average diameter of the bin. Similar to the number based cumulative distribution function, the volume based cumulative distribution function is calculated as:

$$\text{Volume CDF} = \frac{\sum_{i=1}^{i=j}\left(\frac{dV}{d(\log D)}\right)_i \delta(\log D)_i}{\sum_{i=1}^{i=i_\max}\left(\frac{dV}{d(\log D)}\right)_i \delta(\log D)_i} \quad (3)$$

Both number and volume based size distribution functions are examined here since they are the distribution functions most commonly reported via PDPA and LD, respectively. Hence, we demonstrate that DIH enables determination of both.

Droplet asphericity typically correlates with droplet size, and an eccentricity PDF does not fully describe droplet dynamics in a spray process. Two-dimensional joint PDFs have been used with increasing frequency in aerosol analysis (Broda *et al.*, 2018; Chen *et al.*, 2018) to examine the structures of non-spherical submicrometer particles. We extend this principle here by constructing a two-dimensional, volume based size-eccentricity joint PDF. For this purpose one-dimensional volume based size distributions are further binned into $j_{max}$=100 linearly spaced eccentricity bins, yielding $\left(\frac{dV}{d(\log D)de}\right)_{i,j}$ the volume measured per unit log change in diameter and per unit change in eccentricity, for the $i^{th}$ diameter bin and $j^{th}$ eccentricity bin. The joint PDF is then computed as:

$$\text{Joint PDF}_{i,j} = \frac{\left(\frac{dV}{d(\log D)de}\right)_{i,j} \delta(\log D)_i \delta e_j}{\sum_{j=1}^{j=j_{max}} \sum_{i=1}^{i=i_{max}} \left(\frac{dV}{d(\log D)de}\right)_{i,j} \delta(\log D)_i \delta e_j} \qquad (4)$$

*3.2 Monodisperse Droplet Measurements*

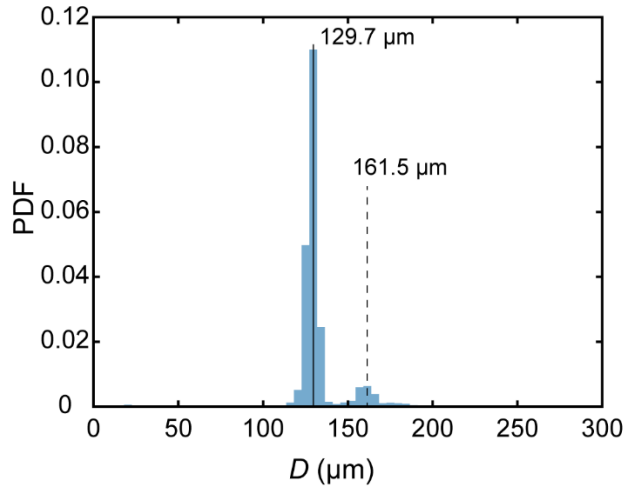

**Figure 4.** The number based probability distribution function determined using DIH for droplets produced by a vibrating orifice aerosol generator.

Figure 4 displays a plot of the DIH determined PDF for VOAG generated droplets. Approximately 40,000 droplets were used for PDF construction. The main peak appears at 129.7 μm which falls within ~0.85% of the nominal droplet diameter calculated for the VOAG settings used (128.6 μm), demonstrating the accurate sizing capability of DIH. Additionally, the peak is narrow in width with a full width at half maximum (FWHM) of ~9.1 μm (and a geometric standard near 1.03). Assuming the VOAG generated initial droplets are perfectly monodisperse, this gives an effective sizing resolution for DIH of 14.2, where the resolution is defined as the ratio of peak location to FWHM of the signal. This resolution is a function of size (it improves with increasing size), and can be further improved through adjustments to the resolution of the holograms themselves. Interestingly, DIH measurements also reveal a second, less pronounced peak at 161.5 μm, which we believe to correspond to the formation of dimers (i.e., a droplets of double the volume). The theoretical dimer size of 163.4 μm, estimated as $2^{1/3}$ times the main peak, deviates only 1.1% from the observed value here. Though dimer formation was unintentional, it is beneficial as they serve as an additional means to confirm the accurate sizing ability of DIH.

## 3.3 Polydisperse Flat Fan Spray Droplet Measurements

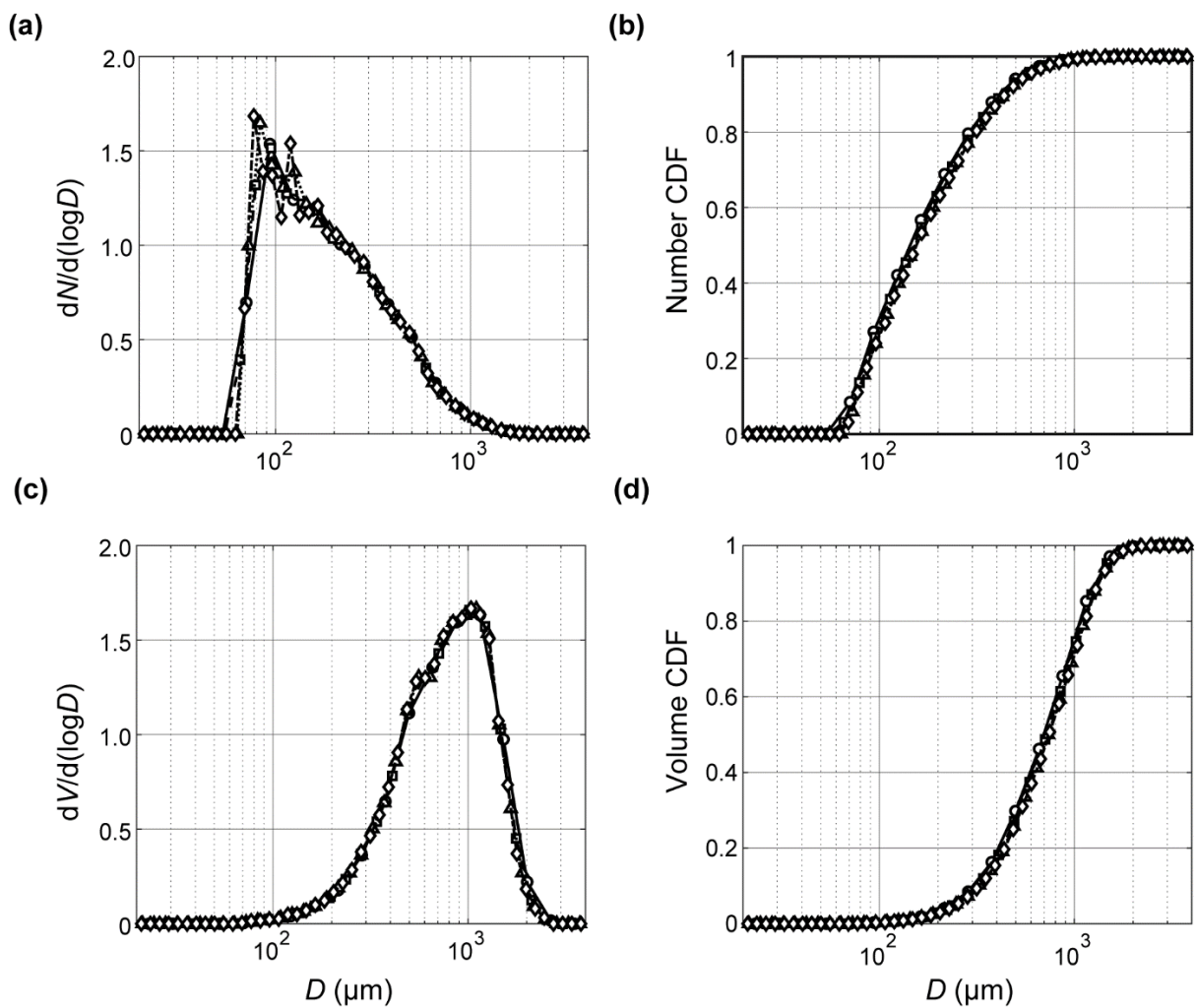

**Figure 5.** Size distribution functions for XR6515 flat fan nozzle spray generated droplets at a distance 304.8 mm (74.3$D_N$) downstream of the nozzle (position 1). **(a)** The number based size distribution function; **(b)** the number based cumulative distribution function; **(c)** the volume based size distribution function; and **(d)** the volume based cumulative distribution function. Each figure plots distribution functions with a variable number of bins: 20-circles, 30-squares, 40-triangles, and 50-diamonds.

Droplet size distribution functions on both a number and volume basis, as well as their corresponding cumulative distribution functions, are shown in Figure 5 for XR6515 flat fan nozzle spray generated droplets at position 1, 304.8 mm (74.3$D_N$) downstream from the nozzle. All distribution functions are constructed using measurements for ~1.6 × 10$^6$ droplets and are plotted using different numbers of bins as discussed in *Section 3.1*. Evident in all plots, the number of bins utilized minimally influences results in the range examined. As expected for flat fan sprays, droplets are highly polydisperse (Kooij *et al.*, 2018) and fitting to a lognormal model we find geometric mean diameters of 145.3 μm and 806.6 μm with geometric standard deviations of 1.92 and 1.77 on a number and volume basis, respectively (curve fits are shown in comparison to measurements in the supporting information). We remark that the truncation at the smallest diameters in the number based size distribution function is likely the result of the dimensions of droplets approaching the imaging resolution of the current DIH system. As such the resolution at these diameters can be readily improved using imaging lens of higher magnification and numerical aperture.

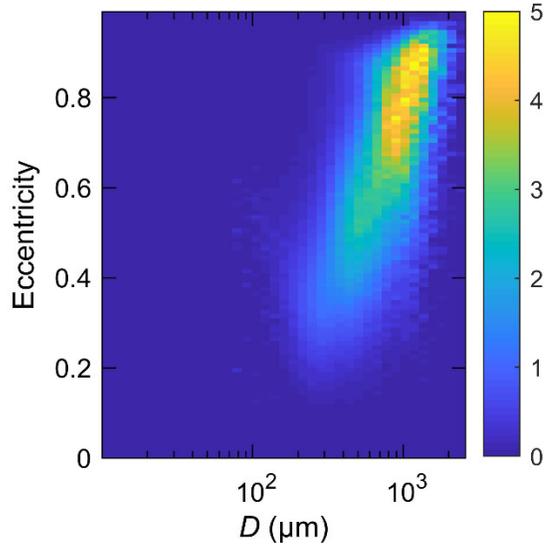

**Figure 6.** The volumetric size-eccentricity joint PDF for measurements at a distance of 304.8 mm (74.3$D_N$) downstream of the XR6515 flat fan nozzle (position 1).

Figure 6 displays a contour plot of the volumetric size-eccentricity joint PDF measured at position 1, which indicates a clear semilogarithmic scaling between the eccentricity and the diameter, reaching a value of ~0.8 for droplets of 1 mm diameter. The spray process typically creates droplets through pinch off from ligaments, resulting in drop oscillations, which are ultimately damped out through viscous dissipation. As the dissipation scales inversely with size, droplets of smaller diameters often relax to a spherical shape much faster than the larger ones in a polydisperse spray. Our results highlight this difference in eccentricity for droplets imaged at a fixed distance from the nozzle, where we can expect to measure (on average) a higher eccentricity for larger diameters. Such results cannot be obtained with alternative size distribution measurement techniques, and moreover, the finding of such highly nonspherical droplets suggests that errors in sizing can manifest with LD and PDPA in size distribution inference for flat fan spray droplets. The magnitude of such errors, however, is not clear, and would need to be directly addressed in future work comparing, LD, PDPA, and DIH.

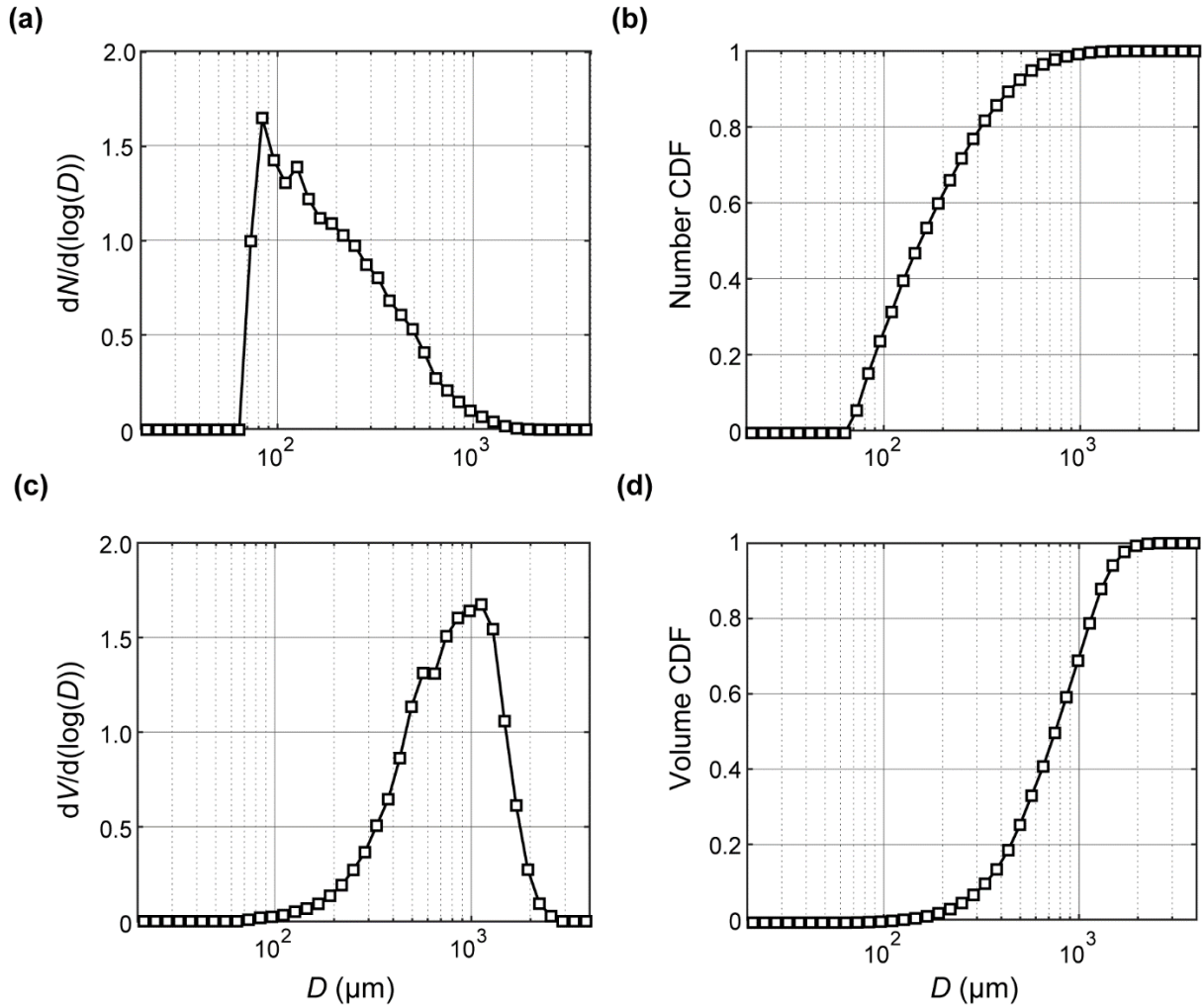

**Figure 7**. Size distribution functions for XR6515 flat fan nozzle spray generated droplets at a distance 457.2 mm (111.5$D_N$) downstream of the nozzle (position 2). **(a)** The number based size distribution function; **(b)** the number based cumulative distribution function; **(c)** the volume based size distribution function; and **(d)** the volume based cumulative distribution function. Each figure plots distribution functions with 30 bins.

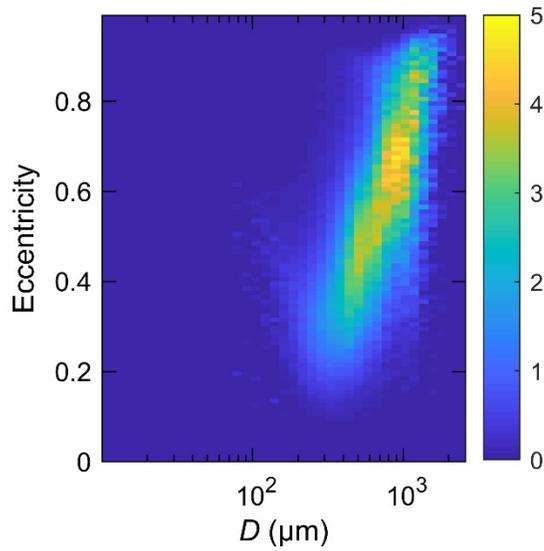

**Figure 8.** The volumetric size-eccentricity joint PDF for measurements at a distance 457.2 mm (111.5$D_N$) downstream of the XR6515 flat fan nozzle.

Figures 7 and 8 show the one dimensional distribution functions and volumetric size-eccentricity joint PDF, respectively, resulting from DIH measurements at position 2. The size distribution functions are very similar at both positions, with geometric mean diameters of 150.9 μm (geometric standard deviation 1.95) in number and 743.6 μm (geometric standard deviation 1.72) in volume. Note the variation in the mean diameters from position 1 to position 2 are ~3.8% and ~7.8% for the number and volume based measurements, respectively, and the good agreement between them confirms the reliability and robustness of DIH for polydisperse measurements. The volumetric joint PDF reveals a semilogarithmic scaling similar to position 1, but with a peak shift towards smaller diameters and eccentricities. As discussed earlier, the relaxation of droplet oscillations would cause this downward shift, which we plan to quantify in the analysis that follows.

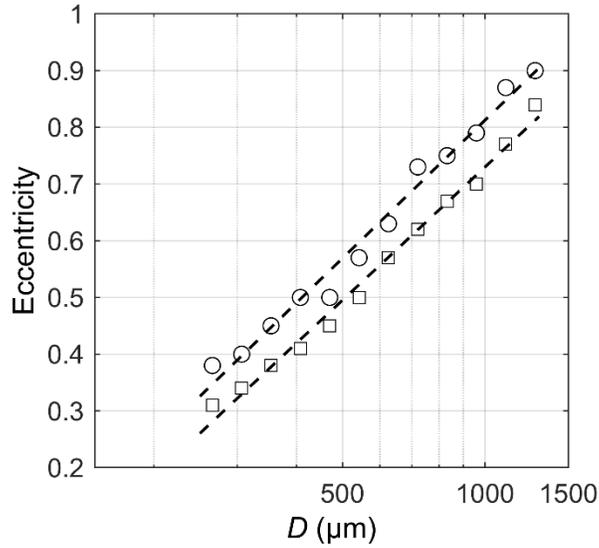

**Figure 9.** A comparison of the mode eccentricity as a function of droplet diameter (*D*) for flat fan spray generated droplets at both measurement positions. Circles- Position 1; Squares- Position 2.

To compare the scaling between the eccentricity and diameter with distance from the nozzle, we extract the eccentricities with the highest probability at each diameter for the two data sets. We present these values along with a semilogarithmic fit to the data in Figure 9. As the figure shows, both data sets yield very similar slopes with a clear vertical offset (of ~0.1) for position 2 relative to position 1. This drop in the peak eccentricities with downstream distance points to an overall relaxation in droplet oscillations across all diameters. Furthermore, we can extrapolate both curves to estimate the droplet size corresponding to zero eccentricity ($D_0$), when droplet oscillations have completely died out, to be 117.0 μm and 138.0 μm at position 1 and 2, respectively. However, we remark that the number of droplets we capture at these sizes are a much smaller fraction of the total volume and thus involves a higher level of uncertainty in the exact count (hence we do not observe the zero eccentricity droplets directly). We anticipate that the eccentricity of all droplets will decrease as they move further away from the nozzle. Finally, it is also worth pointing out that the uncertainty in the eccentricity characterization increases as the droplet size approach the imaging resolution of the current DIH system.

## 4. Conclusions

We have utilized digital inline holography (DIH) for automated droplet size distribution measurements of both vibrating orifice generated (VOAG) and flat fan nozzle sprayed droplets. To our knowledge, this is the first application of DIH for characterizing the two dimensional distribution function of spray droplets. Based on these measurements, we draw the following conclusions:

1. DIH size information is in excellent agreement with the expected size for VOAG generated droplets, both monomers and dimers. Importantly, DIH imaging system calibration can be carried out without the need for monodisperse standard particles. Even with the rather coarse pixel resolution utilized (18.2 μm/pixel), DIH has a size resolution near 130 μm of 14.2. This suggests that DIH may find application in standards characterization and in protocols where spray size distribution functions need to be measured.
2. Converged polydisperse distribution functions can be measured using DIH with size binning carried out as part of post-processing, instead of a priori with predefined bins. Because DIH is an imaging based technique, no prescribed functional forms are needed to fit size distributions, and multimodal distributions can be examined.
3. Flat fan generated droplets are highly aspherical at the point of generation, in contrast with the conventional assumptions made in LD and PDPA operation. DIH shows clearly that the droplet eccentricity scales with droplet size semilogarithmically, and that the eccentricity decreases farther from the spray nozzle. More refined data analysis with time-resolved DIH measurements capturing droplet deformation dynamics can aid in in situ measurement of liquid interfacial tension, an approach that is widely used in microfluidic platforms for small scale droplets (Brosseau *et al.*, 2014).

In total, our current study introduces a fully automated digital inline holography imaging system for characterizing both monodisperse and polydisperse sprays. Apart from providing measurements on a single droplet basis, the complexity of the system in comparison to LD and PDPA is significantly reduced. Furthermore, the cost of a typical DIH system is an order of magnitude smaller than either of the two instruments. For instance, a DIH system can be constructed using a consumer camera, a diode laser module and some other simple optical components with total budget below or near $1000, while conventional LD and PDPA systems costs often well above $10,000. Recent work has also demonstrated the application of DIH in aquatic field settings (Watson, 2011), and the setup is amenable to aircraft and UAV measurements (Beals *et al.*, 2015). We hence regard DIH as a promising technique for aerosol analysis both in the laboratory and in the field. Our measurements have highlighted the high resolution and versatility through DIH imaging of monodisperse droplets from a VOAG as well as polydisperse sprays from a flat fan nozzle, respectively. The technique can provide modest to high resolution (ratio of peak to FWHM >10, with much larger values possible), that is capable of resolving monodisperse and polydisperse distributions, with the simultaneous ability to quantify the extent of asphericity in non-spherical particle and droplets. While our focus in this study was on size and shape distribution functions, the ability to extract multidimensional distribution functions can be extended to examine velocity/flux distributions (Damay *et al.,* 2009) turbulent deposition and dispersion of aerosols (Nicholson, 1988) which we hope to target in future studies.

## 5. Supporting Information

A video resulting from DIH measurements of monodisperse VOAG droplets, a summary of lognormal distribution fitting, notes on calibration using a precision microruler, and additional images from image processing of polydisperse flat fan nozzle generated droplets are available online.

## 6. Acknowledgements

This work was supported by Winfield United. The authors acknowledge the Minnesota Supercomputing Institute (MSI) at the University of Minnesota for providing resources that contributed to the research results reported within this paper. URL: http://www.msi.umn.edu. The authors also thank Dr. Bernard A. Olson and Mr. Ian Marabella (University of Minnesota) for their assistance on using the vibrating orifice aerosol generator for monodisperse measurements.